\documentclass[%
 aps,
 twocolumn,
superscriptaddress,
 amsmath,amssymb,
aps,
prb,
]{revtex4-2}

\usepackage{graphicx}% Include figure files
\usepackage{appendix}
\usepackage{dcolumn}% Align table columns on decimal point
\usepackage{bm}% bold math
\usepackage{hyperref}% add hypertext capabilities
\usepackage{siunitx}
\usepackage{array}
\usepackage{multirow}
\usepackage{float}
\usepackage{chemformula}
\usepackage{tabularx}
\usepackage{textcomp}
\usepackage{subcaption}
\usepackage{physics}
\usepackage{placeins}
\usepackage{array}
\newcolumntype{Z}{>{\raggedright\let\newline\\\arraybackslash\hspace{0pt}}X}

\newcommand{\pr}[1]{\left(#1 \right)} %for () that scale
\renewcommand{\v}[1]{\ensuremath{\mathbf{#1}}} % for vectors
\newcommand{\gv}[1]{\ensuremath{\pmb{#1}}} % for vectors of Greek letters
\begin{document}

\preprint{APS/123-QED}

\title{Prediction of Large Magnetic Moment Materials With Graph Neural Networks and Random Forests}

\author{Sékou-Oumar Kaba}
\email{kabaseko@mila.quebec}
\affiliation{Mila - Quebec Artificial Intelligence Institute \& IVADO - Institut de Valorisation des Donn\'ees, Montr\'eal, Qu\'ebec, Canada H2S 3H1}
\affiliation{School of Computer Science, McGill Univeristy, Montr\'eal, Qu\'ebec, Canada H3A 0E9}
\author{Benjamin Groleau-Paré}
\affiliation{D\'epartement de physique \& Institut Quantique, Universit\'e de Sherbrooke, Qu\'ebec, Canada  J1K 2R1}
\author{Marc-Antoine Gauthier}
\affiliation{D\'epartement de physique \& Institut Quantique, Universit\'e de Sherbrooke, Qu\'ebec, Canada  J1K 2R1}
\author{A.-M.S. Tremblay}
\affiliation{D\'epartement de physique \& Institut Quantique, Universit\'e de Sherbrooke, Qu\'ebec, Canada  J1K 2R1}
\author{Simon Verret}
\affiliation{Mila - Quebec Artificial Intelligence Institute \& IVADO - Institut de Valorisation des Donn\'ees, Montr\'eal, Qu\'ebec, Canada H2S 3H1}
\affiliation{D\'epartement de physique \& Institut Quantique, Universit\'e de Sherbrooke, Qu\'ebec, Canada  J1K 2R1}
\author{Chloé Gauvin-Ndiaye}
\affiliation{D\'epartement de physique \& Institut Quantique, Universit\'e de Sherbrooke, Qu\'ebec, Canada  J1K 2R1}

\date{\today}% It is always \today, today,
             %  but any date may be explicitly specified

\begin{abstract}
Magnetic materials are crucial components of many technologies that could drive the ecological transition, including electric motors, wind turbine generators and magnetic refrigeration systems. Discovering materials with large magnetic moments is therefore an increasing priority. Here, using state-of-the-art machine learning methods, we scan the Inorganic Crystal Structure Database (ICSD) of hundreds of thousands of existing materials to find those that are ferromagnetic and have large magnetic moments. Crystal graph convolutional neural networks (CGCNN), materials graph network (MEGNet) and random forests are trained on the Materials Project database that contains the results of high-throughput DFT predictions. For random forests, we use a stochastic method to select nearly one hundred relevant descriptors based on chemical composition and crystal structure. This gives results that are comparable to those of neural networks. Our findings suggests that magnetic properties are intrinsically more difficult to predict than other DFT-calculated properties. The comparison between the different machine learning approaches gives an estimate of the errors for our predictions on the ICSD database. Validating our final predictions by comparisons with available experimental data, we found 15 materials that are likely to have large magnetic moments and have not been yet studied experimentally. 
\end{abstract}

%\keywords{Suggested keywords}%Use showkeys class option if keyword
                              %display desired
\maketitle

%\tableofcontents

\section{Introduction}

% ML for material discovery
In recent years, materials informatics and machine learning methods have been introduced in the search for materials with specific properties, such as high-temperature superconductors \cite{doi:10.1063/5.0004641}, photovoltaics \cite{ref-photovoltaics}, radiation detector materials \cite{ortiz2009} and metallic glasses \cite{ward2016}. These methods have the advantage of allowing to explore sets of materials that would be prohibitively large for conventional theoretical methods or experiments. Though traditional machine learning methods such as tree-based algorithms \cite{landrum2003, rhone2020},  kernel methods \cite{moller2018, rhone2020}, support vector machines \cite{rhone2020} and multilayer perceptrons \cite{rhone2020} have shown some success in the prediction of magnetic properties, the frequent introduction of new large materials databases \cite{ref-materialsproject, oqmd, magnetic_database, aflow} has also enabled the development of sophisticated neural networks for this type of application. In particular, recent Graph Neural Networks algorithms (GNNs) have recently been shown to obtain state-of-the-art performance on benchmark tasks \cite{cgcnn, megnet, schutt2017schnet}. These methods have been shown to successfully predict formation energy, band gap and bulk modulus with error magnitudes similar to those of density functional theory (DFT) calculations, but have not yet been used for the study of magnetic properties. %Considering the challenges that surround the design of new materials for magnetic refrigeration, rotative or conventional, we investigate the usefulness of machine learning for materials in the study of magnetic properties as a first step towards the discovery of new magnetocaloric materials.

Demand for strong permanent magnets for technological applications is rising \cite{nakamura2018}. This is closely related to the fact that many countries are looking to transition away from fossil fuels to more sustainable energy sources. Indeed, one of the main drivers of demand for permanent magnets is the production of motors for hybrid and electric vehicles, which are rapidly gaining popularity. Another growing application is wind turbine generators. For most applications, Nd$_2$Fe$_{14}$B is the material of choice. However, discovering rare-earth free permanent magnets would be highly desirable for environmental and economic reasons. 

Materials that have a large magnetic moment per mass unit, but that are not permanent magnets, still have multiple applications of interest. A promising application for such materials is magnetic refrigeration. Magnetic refrigeration is a technology based on the magnetocaloric effect, through which the temperature of a magnetic material varies with the adiabatic application of a magnetic field \cite{ref-weissmce, ref-mce, doi:10.1002/aenm.201200167}.  Because it requires the use of solid state materials instead of gaseous refrigerants, magnetic refrigeration is a more environmentally-friendly technology than traditional refrigeration.  However, it requires ferromagnetic materials with a Curie temperature around room temperature that also have many specific properties, such as a low specific heat and a high electrical resistivity. The reference materials that exhibit a large magnetocaloric effect around room temperature are gadolinium and some Gd-based alloys like Gd$_5$Si$_2$Ge$_2$ \cite{brown_gd, ref-gd}. These materials are expensive, which limits their commercial use, and are metallic, which diminishes energy efficiency due to heat loss.

Often, the search for new materials that exhibit specific properties is done through trial and error, limiting the number of materials that can be studied both theoretically and experimentally. Considering the challenges that surround the design of new materials for magnetic refrigeration and other applications, we investigate the usefulness of machine learning for materials in the study of magnetic properties as a first step towards the discovery of new materials.

\begin{figure*}
         \centering
         \includegraphics[width=0.8\textwidth]{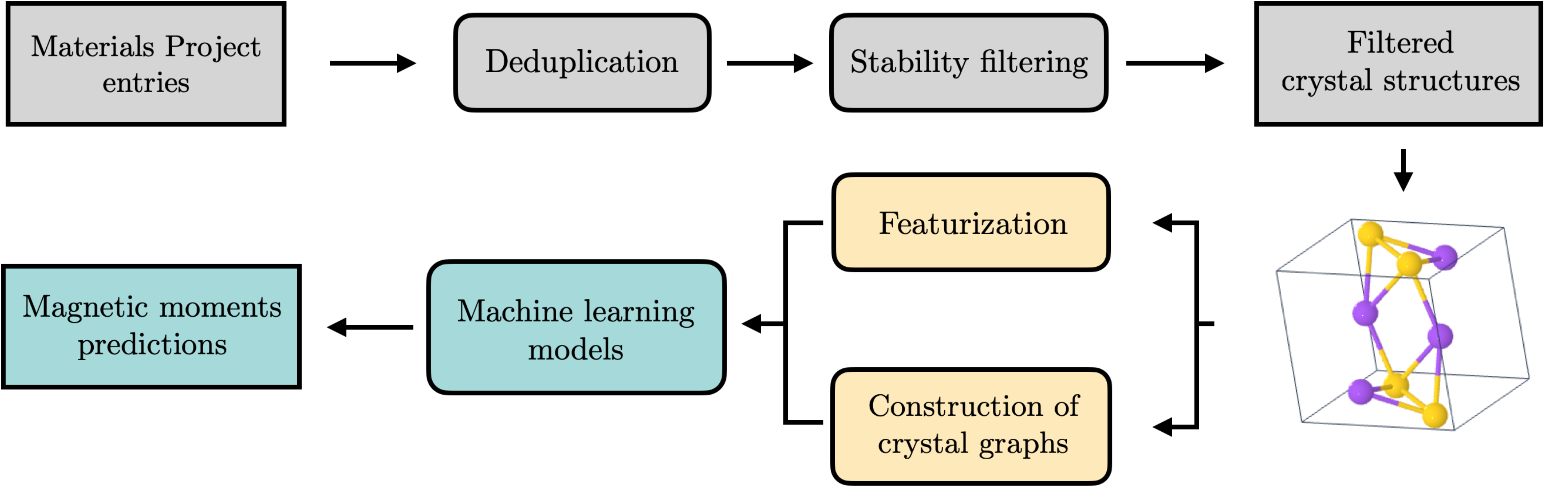}
        \caption{Data-flow for model training, including cleaning, preprocessing, and prediction steps.}
        \label{fig:dataflow}
\end{figure*}

% Plan of the paper
In this paper, we assess the performance of two recently proposed neural networks, CGCNN \cite{cgcnn} and MEGNet \cite{megnet}, and compare it to that of the random forest, a statistical machine learning method~\cite{breiman2001random,Intro_to_stat,liaw2002classification}, for the prediction of the magnetization of materials. To do so, we train our models on the Materials Project database, a DFT database frequently used as a training set in the field of machine learning for material properties \cite{ref-materialsproject}. We characterize the Materials Project dataset and describe a preprocessing scheme based on the energy above hull to reduce bias in its distribution of magnetic orders. Figure~\ref{fig:dataflow} shows the steps used in training the models. Our work reveals that neural networks are comparable to random forests for the prediction of magnetization. We also find that the performance of machine learning models in fitting magnetization is comparatively worse than for other properties like formation energy, suggesting that it is an intrinsically more difficult task. We then apply our trained models on the ICSD database, which contains around $100~000$ stochiometric experimentally-studied materials \cite{ref-icsd}. We discuss the suitability of the proposed materials for the specific application to magnetic refrigeration.

In the following section, we describe the datasets. Sec.~\ref{Sec:Methods} explains the machine-learning methods, followed by our predictions and error estimates in Sec.~\ref{Sec:Results}. A discussion in Sec.~\ref{Sec:Discussion} is followed by the conclusion in Sec.~\ref{Sec:Conclusion}.

\section{Datasets}
We first discuss the Material's Project database that was used to train the models and then the ICSD database of materials that we use to make predictions. 

\subsection{Materials Project}

% background information : history, version used, number of compounds

The training data used in this work comes from the Materials Project dataset (V2020.06) \cite{ref-materialsproject}. It is one of the largest datasets obtained from high-throughput DFT calculations and has become standard in machine learning based materials studies.

Materials Project comprises stochiometric crystalline materials, and provides their chemical composition, relaxed structures and a number of properties such as the formation energy, the energy above hull, the band structure and the spontaneous unit cell magnetization. To our knowledge, this last property in Materials Project has not been used yet for machine learning applications. Our first objective is to determine whether or not the magnetization is a property that can be modeled properly with machine learning algorithms.

% dft algorithm used : advantages, drawbacks for magnetization
We note that all the calculations are initialized in the ferromagnetic configuration. Antiferromagnetic configurations can be reached in the crystal relaxation stage. However, it has been shown that this method favours ferromagnetic configurations, even when low-spin antiferromagnetic or ferrimagnetic configurations could have lower energies \cite{afm-mp}. In 2019, a new workflow was introduced as an effort to include appropriate antiferromagnetic ground states and counter the ferromagnetic bias \cite{afm-mp}. At this time, new ground-state calculations using this workflow were performed for about $520$ materials, less than 2\% of the Materials Project dataset. 

The most difficult materials to simulate with DFT calculations are the ones in which electronic correlations are strong \cite{Pavarini_2021}. This is most notable in materials that contain $d$ or $f$ valence electrons, like the transition metals and the rare earths. These electrons are also the ones that participate in the magnetic properties. In the Materials Project dataset, all calculations on oxides containing Co, Cr, Fe, Mn, Mo, Ni, V and W atoms are performed with the GGA+U scheme that aims to better represent the electronic correlations. We still note that the magnetic ground state can be significantly influenced by the effect of electron-electron interactions beyond the GGA+U approximation.

We filter the entries of the Materials Project dataset to create our training set. This is because the predictions of our machine learning models can only come from the identification of patterns within this dataset. It is therefore crucial to identify biases and potential obstacles to generalization in this training distribution \cite{gudivada2017data}. We first remove duplicates from the dataset. Indeed, in the dataset there is an entry for each calculation, and not each material, which results in having entries with only marginal structural differences. We use a simple heuristic to remove duplicates from the dataset. Two materials are considered similar if they share the same unit cell composition and space group. This criterion is conservative, but computationally tractable and allows to effectively remove duplicates.

\begin{figure}[h]
\includegraphics[width=1\columnwidth]{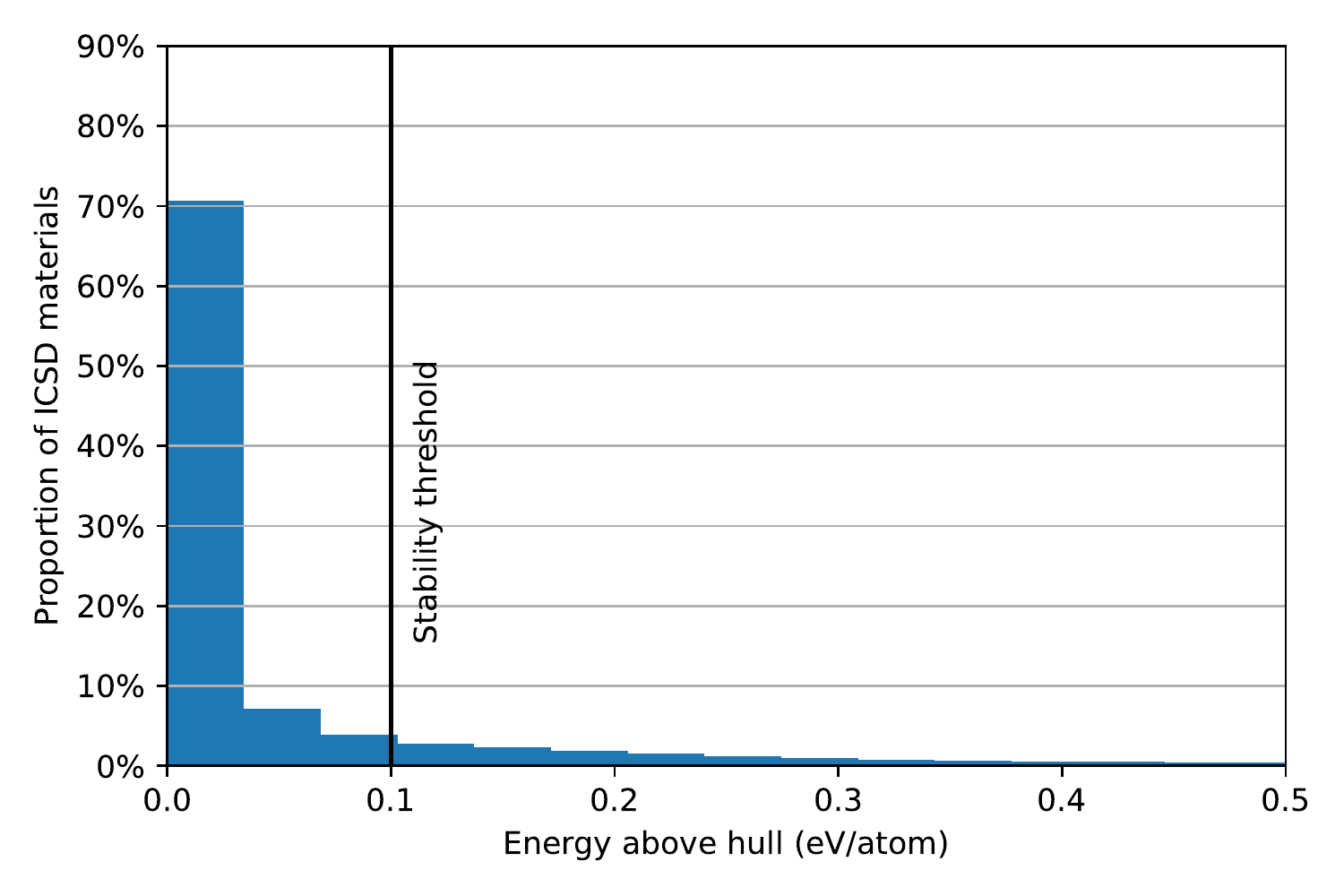}% Here is how to import EPS art
\caption{\label{fig:eabovehull_icsd} Distribution of the energy above hull attribute in the Materials Project dataset for the subset of entries that are also in the ICSD dataset.}
\end{figure}

\begin{figure}[h]
\includegraphics[width=1\columnwidth]{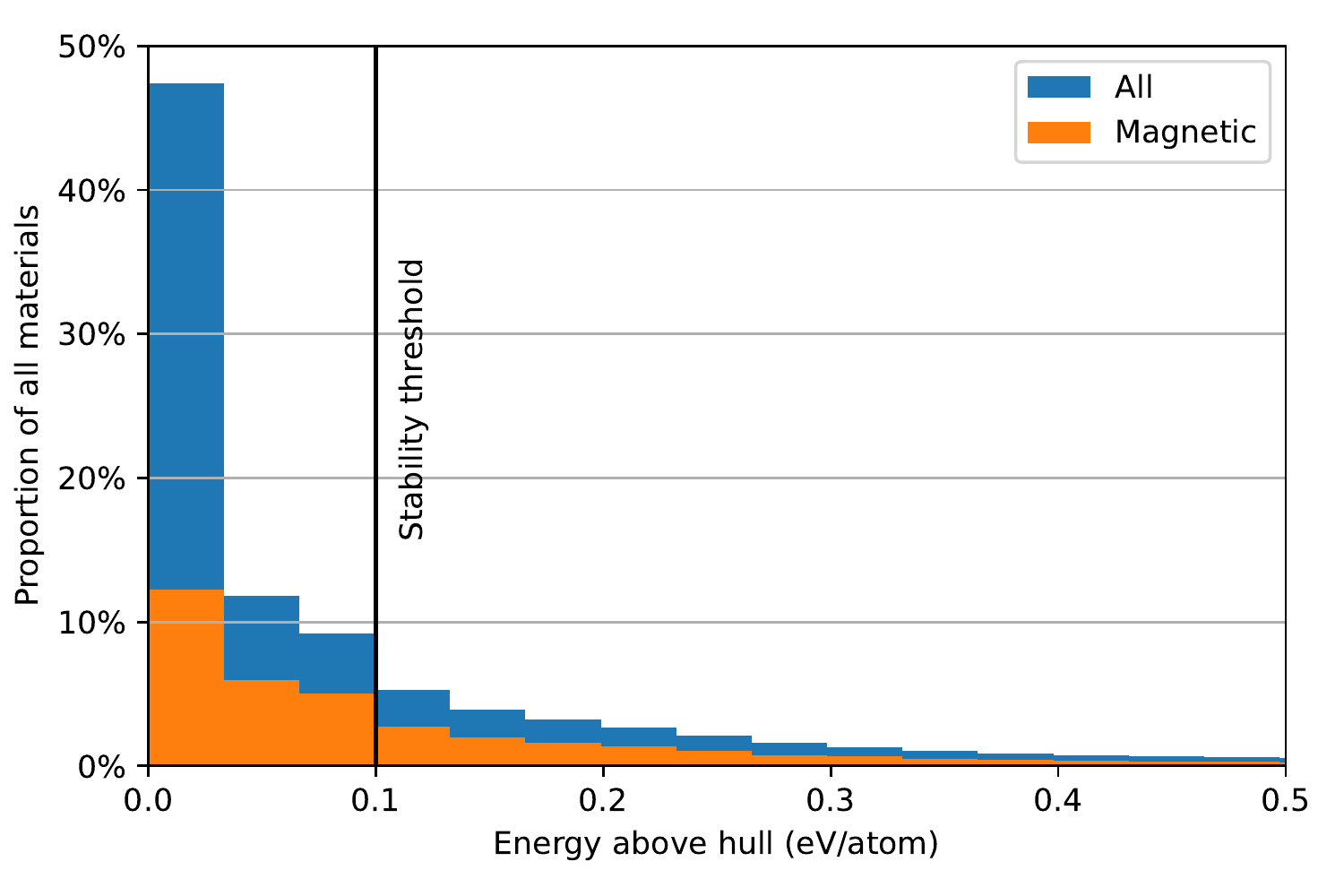}% Here is how to import EPS art
\caption{\label{fig:eabovehull} Distribution of the energy above hull attribute in the Materials Project dataset.}
\end{figure}

Another important source of errors comes from the fact that DFT can relax to unstable, high-energy structures. Not only these materials cannot be synthesized and will drive the distribution away from materials of interest, but they will also tend to exhibit atypical features that could confuse training. We thus choose to filter these materials out of the dataset based on their energy above hull computed by DFT. The energy above hull gives the energy of decomposition of a material into a set of stable materials containing all the chemical elements of the original material and whose total formation energy is smaller. As shown in Figure \ref{fig:eabovehull_icsd}, the distribution of energies above hull is highly skewed towards small values for entries that also appear in ICSD (i.e. materials that have been synthesized). We choose heuristically the value of $E = 0.1 \text{eV}/\text{atom}$ as a threshold for stability. This results in keeping $69\%$ of the materials in the dataset (Figure~\ref{fig:eabovehull}).
The total curated dataset leaves us with 78~462 materials.

The Materials Project dataset is not specifically focused on magnetic materials and includes non-magnetic (which includes paramagnetic since only spontaneous magnetization is reported), ferromagnetic and antiferromagnetic materials. For our goal of predicting magnetization, having both non-magnetic and magnetic materials is a desirable property: we want the model to identify the factors promoting strong magnetization as well as those inhibiting it. Figure~\ref{fig:eabovehull} shows that the proportion of magnetic materials increases with the energy above hull, so eliminating large energy above hull material helps fill this criterion of having a more balanced proportion of magnetic and non-magnetic materials.

\begin{figure}[h]
\includegraphics[width=1\columnwidth]{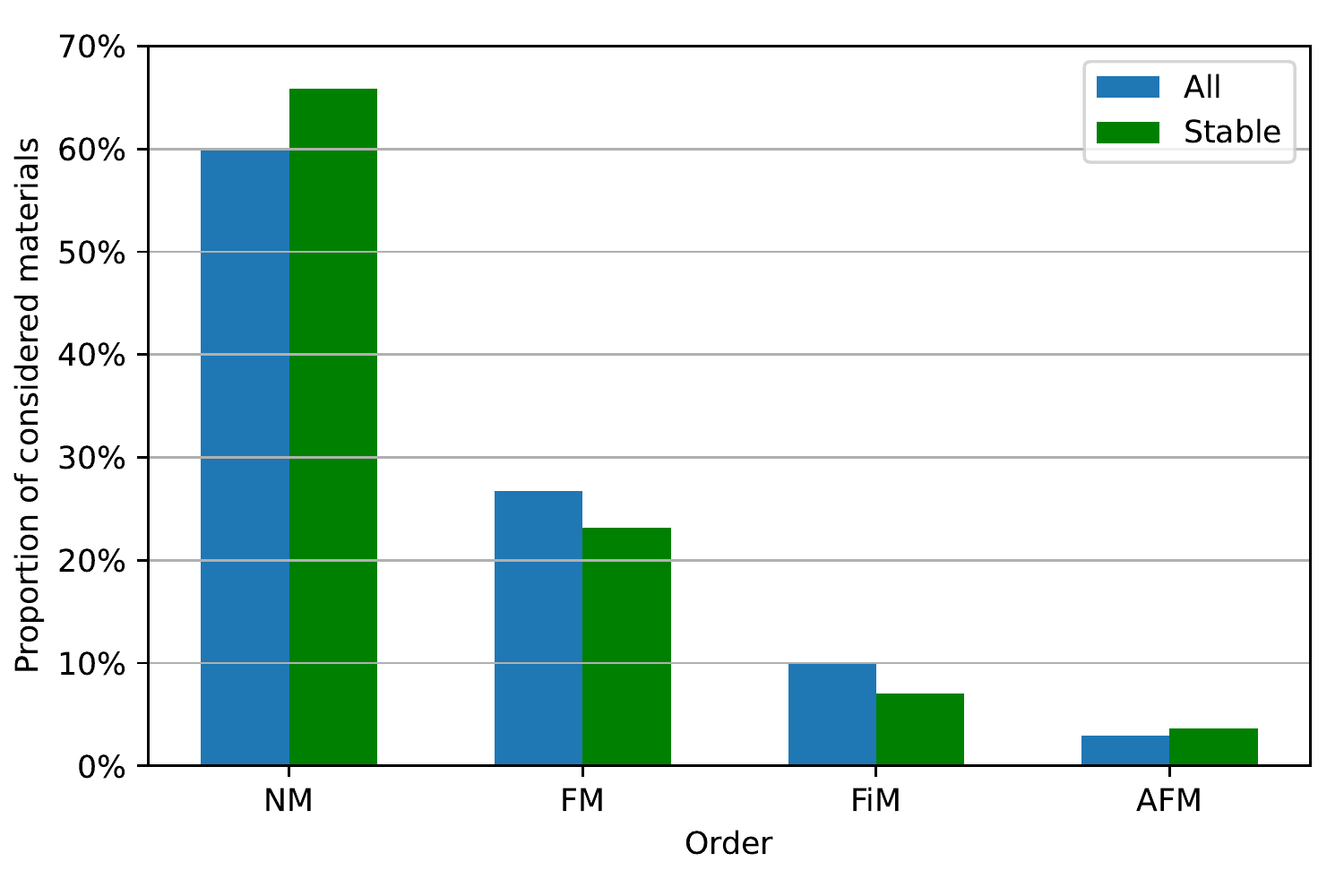}% Here is how to import EPS art
\caption{\label{fig:orderings} Distribution of magnetic orders in the Materials Project dataset. NM stands for non-magnetic, FM for ferromagnetic, FiM for ferrimagnetic and AFM for antiferromagnetic.}
\end{figure}

The number of materials for each magnetic order is indicated in Figure~\ref{fig:orderings} for all materials and for those that are stable according to our criterion. We note that the procedure of filtering out unstable structures helps to alleviate the ferromagnetic bias of DFT.  Figure~\ref{fig:orderings} further shows that the eliminated materials are more ferromagnetic and ferrimagnetic; the proportion of antiferromagnetic materials slightly increases after filtering.

Further details on Materials Project are provided in Appendix A of the Suplementary Material.

\subsection{ICSD}

Having trained models on the Materials Project, our second objective is to identify high magnetization candidates that can readily be synthesized. We use data from the Inorganic Crystal Structures Database (ICSD) to perform this task. It is currently the largest database of experimentally identified crystalline materials.

ISCD data is obtained directly from scientific publications. It includes chemical composition as well as crystal structure data for all of its entries.

Unlike the Materials Project, ICSD includes non-stochiometric materials. Since our training distribution did not include such materials, we expect that our models will have difficulty generalizing to these materials and we have thus removed these entries from the inference dataset. In addition, it is crucial to take into account that the Materials Project and ICSD datasets are not independent. The crystal structures of the entries in Materials project are computed starting from ICSD entries. Though some of the structures change significantly through the relaxation process, many of them remain sufficiently similar to the ICSD starting point to be considered as common entries between the two datasets. We have identified the common entries in both datasets and found that out of the $110~870$ entries in ICSD that are stochiometric and compatible  with the methods, $23~311$ materials did not have matching identifiers or formulas. Inference of magnetization was therefore perfomed on this subset.

\section{Methods}
\label{Sec:Methods}

Once trained, machine learning algorithms have the crucial advantage of producing predictions orders of magnitude faster than simulation methods like DFT. We can thus use them to efficiently screen candidate materials in a large database. The setup is that of a supervised learning task: given a training set of materials with features $\v{X}$ and known ground-truth targets $\v{Y}$, the algorithm is tasked to learn a prediction function $\v{\hat{Y}} = f_{\gv{\theta}}\pr{\v{X}}$. Training parameters $\gv{\theta}$ are optimized to minimize a loss function $\mathcal{L}\pr{\v{Y},\v{\hat{Y}}}$. 

\begin{figure*}
     %\centering
    %  \begin{subfigure}{\textwidth}
    %      \centering
         \includegraphics[width=0.8\textwidth]{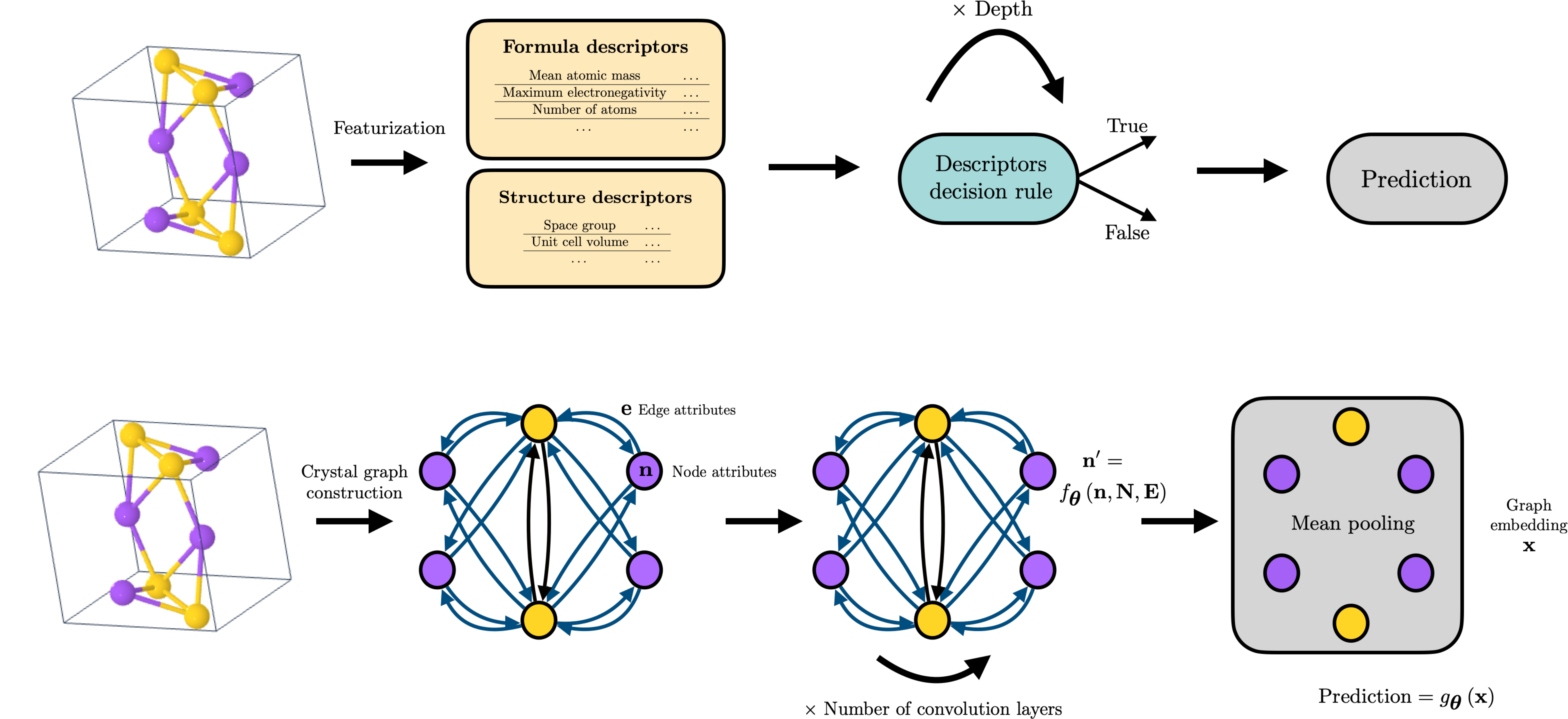}
    %      \caption{}
    %      \label{fig:tree}
    %  \end{subfigure}
    %  \hfill
    %  \begin{subfigure}{\textwidth}
    %      \centering
    %      \includegraphics[width=0.9\textwidth]{figures/figure_gnn.pdf}
    %      \caption{}
    %      \label{fig:gnn}
    %  \end{subfigure}
        \caption{Methods used for property prediction. Top panel: Illustration of a decision tree. First, descriptors are computed from the chemical formula and the structural properties. A series of decisions are then taken using the descriptors. After a number of decisions, the final prediction is made. Bottom panel: Illustration of a GNN. A graph is built from the crystal unit cell. Each atom is mapped to a node and an edge is drawn between two nodes if they share a Voronoi face. Both nodes and edges have associated embedding vectors. The graph then goes through a series of graph convolution operations (defined differently for each method) parameterized by the neural network $f_{\gv{\theta}}$. Finally, the features of all the nodes are averaged and the resulting vector goes through a multilayer perceptron $g_{\gv{\theta}}$ that outputs the prediction.}
         \label{fig:models}
        \label{fig:methods}
\end{figure*}

In the following subsections, we describe the learning algorithms we have used. Hyperparameters are given in Appendix B of the Supplementary Material. Aside from the more complex machine learning methods discussed in this paper, we also used the linear model~\cite{Scikit}. The linear model is trained using the same descriptor space as the random forest. Since the linear model assumes a simple relation between each descriptor and the target, it is not expected to be accurate. Therefore, it establishes a performance baseline against which more elaborate methods will be compared. Figure~\ref{fig:methods} summarizes schematically the methods that we use. We explain each of them in more detail below.

\subsection{Random Forests}

The random forest \cite{random_forest} is a tree-based machine learning algorithm that has been widely used for materials property prediction \cite{ward2016, landrum2003, rhone2020}. Random forests have yielded encouraging results for similar material design tasks, for example finding superconductors~\cite{ML_sup, iMat}.
At the root of any tree-based method lies the decision tree, which carries out consecutive binary splits in the descriptor space of the data (see Figure \ref{fig:models}). Single decision trees, however, have the major drawback of frequently overfitting the training data. Random forests go around this problem by averaging the predictions of multiple decision trees. Each tree is built differently to ensure that there is some variance in the generated forest. Randomness is implemented by choosing a subset of the descriptors to be available to the usual tree algorithm every time the algorithm makes a split. The subset is different for every split. The number of descriptors in the subset is a hyperparameter called ``Available features per split". The use of random forests is motivated in our case by their relative simplicity and efficiency \cite{Understanding_rfs}. We use \textsc{scikit-learn}'s \cite{Scikit} implementation of the random forest methods.

The model takes descriptors handcrafted for each sample material as input. The algorithm achieves much better performance if descriptors are well adapted to the task. The properties found in both Materials Project and ICSD, such as density and crystal structure, can be employed in our case. Inspired by other works \cite{iMat, ML_sup}, most of the descriptors are obtained from the chemical formula of the materials: starting from atomic properties, such as the ground-state magnetic moment, the electronegativity, the atomic mass or the ground state d-shell electrons, we compute the mean value, the maximum value or the standard deviation of each of these properties to form descriptors. A number of descriptors are also obtained from the crystal structure, for example the number of sites in the unit cell or bond lengths. We design more than 400 descriptors in this way for each material encountered. Appendix D of the Supplementary Material gives details on descriptor design and provides an exhaustive list of all descriptors available to our random forest model. 

Using this large descriptor space would lead to overfitting. Hence we identify a subset of descriptors that, in addition to yielding better predictions, comes with the added benefit of giving more interpretability to the model.  Because of the large number of descriptors and material entries, forward and backward descriptor selection methods one would typically use for this task require an unreasonable amount of computational resources and time. We therefore design a descriptor selection scheme that mixes both forward and backward descriptor selection in order to efficiently find which descriptors are relevant for the task (details in Appendix E of the Supplementary Material). 

\subsection{Graph Neural Networks}

Graph neural networks (GNNs) are deep learning architectures that are widely used for molecular-property prediction and generation \cite{duvenaud2015, gilmer2017}. Their use in the context of materials is however recent. The main advantage of using GNNs for materials property prediction compared to other machine learning methods is that they only take as input a graph encoding of the material that naturally encodes structure information. The need for using handcrafted features is eliminated, as the deep model acts as a \textit{feature extractor}. This ability to learn a representation adapted to the task at hand from the raw data has been key to the success of deep learning in a variety of domains. However, it is essential to note that this additional expressivity (the complexity and diversity of prediction functions that can be learned) comes at the cost of interpretability. It is notoriously difficult to know how a neural network selects specific features for predictions \cite{chakraborty2017interpretability}. For this reason, we deem more appropriate to use neural networks in conjunction with random forests, a method that allows explicit descriptor construction and selection.

Here we use two architectures, CGCNN \cite{cgcnn} and MEGNet \cite{megnet} that have been chosen for their performance on other properties of the Materials Project dataset as well as their good training speed. For each material, the crystal unit cell is mapped to a sparse graph. This is done by associating each atom to a node and linking two nodes if they share a Voronoi face and are within a cutoff distance of 5 angstroms. Edges are also added if a face is shared with an atom outside the unit cell to enforce periodic boundary conditions. Around 5\% of structures resulted in disconnected graphs and were discarded. When building the graphs, node features are added by taking one-hot encodings of the corresponding atomic type. This allows to capture all the information on atoms composing the unit cell. An encoding based on the distance for edges is used as well. Taking inspiration from Schnet~\cite{schutt2017schnet}, this distance is expanded on a Gaussian basis of functions, with details specified in Appendix B of the Supplementary Material.

Each architecture takes these labeled graphs as input and applies successive graph convolution layers to them (see Figure \ref{fig:models}). These convolution operations consist of each node aggregating the features of its neighbors using a learned function. The main difference between the two models revolves around the design of these layers, which is detailed in the original papers. After a number of passes in convolution layers, node and edge representations are pooled and sent to a regressor multilayer perceptron that outputs the final prediction. For each architecture, we use the original hyperparameters with a few modifications detailed in Appendix B in the Supplementary Material \cite{supplementary-material}. Training is performed using stochastic gradient descent with the \textsc{adam} optimizer \cite{adam} as well as a learning rate scheduler.

\subsection{Training and inference}

We first train a model for each method on the Materials Project dataset. Training is performed by minimizing the mean squared error (MSE) of the predicted magnetization per atom with the Materials Project target
\begin{align}
& \mathcal{L}_{\text{MSE}} = \frac{1}{N} \norm{\v{Y} - \v{\hat{Y}}}_2^2,
\end{align}
where $N$ is the size of the dataset. We also report the mean absolute error (MAE) between predicted magnetization and ground-truth values
\begin{align}
& \mathcal{L}_{\text{MAE}} = \frac{1}{N} \norm{\v{Y} - \v{\hat{Y}}}_1.
\end{align}
The Materials Project dataset is split into a training set comprising 80\% of the data, a validation set of 10\% and a test set of 10\%. We compare the performance on the test of each method in Table~\ref{results_mp}. We also train models to predict the formation energy given in Materials Project. Since results on this task are obtained in the original implementations of CGCNN and MEGNet, this allows us to verify that our implementations of the models perform as well as expected.

Inference is performed using the trained model on the ICSD database. As explained above, non-stochiometric (doped) materials are eliminated since they are absent from the training set. In addition, a small fraction of the entries had incomplete crystal structure information and had to be discarded when using GNNs.

\section{Results}
\label{Sec:Results}

We start by comparing the behaviour of the various methods for the prediction of the magnetization. Then, we give the predictions for magnetic moments of compounds in the ICSD dataset. We also discuss the importance of the random forest descriptors in Appendix F of the Supplementary Material.

\subsection{Evaluation of methods}
We first look at the task of predicting formation energy used as a benchmark for GNN methods. We find that our models perform on par or better than the original implementations. The difference can be attributed to the fact that we use a  different version of the Materials Project dataset as well as different training, validation and test set splits. Our hyperparameters are also slightly different, as detailed in Appendix B of the Supplementary Material \cite{supplementary-material}. We see that neural networks outperform random forests (Table \ref{results_mp_energy}) by a significant margin.

Then, we evaluate the different models on the magnetic moment prediction task on the Materials Project dataset. Results are shown on Table \ref{results_mp}. As expected, all models outperform the baseline linear model. We find that both neural network architectures perform worse on this task than random forests. This is in strong contrast with the evaluation of prediction accuracy of formation energy. Thus, the common assumption that deep models should perform better than models based on handcrafted descriptors does not hold for the task of predicting the magnetic moment.

The difference between magnetization and formation energy results may be understood by the fact that random forests may handle imbalance in prediction labels better than neural networks. In our case, the imbalance is caused by the bimodal distribution of magnetization values, with one mode associated with non-magnetic and antiferromagnetic materials and the other with ferromagnetic and ferrimagnetic materials. These modes have significantly different weights in the distribution as shown in Figure \ref{fig:orderings}. It is well known that neural networks are difficult to train on imbalanced data \cite{johnson2019survey}.
Deep models also show a stronger tendency to overfit training data which can explain that they compare worse on MSE than MAE. This is also confirmed by the MSE on the training set at Table VII (Appendix C of the Supplementary Material), which is one order of magnitude smaller for neural networks than for random forests.

Finally, all methods show only a smaller improvement in performance with respect to the linear model compared to the improvement on formation energy prediction. The reasons for this difficulty of predicting magnetization with more elaborate machine learning are unclear. One could argue that this could be due to the biased distribution of magnetization data in Materials Project. This is however not an entirely satisfying explanation because the test distribution is just as biased as the training distribution is. Our results instead suggest that magnetization is an intrinsically more challenging property to predict than formation energy. Better predictions could then be obtained either by incorporating more physics-based biases into the models or by accumulating more data.

\begin{table}[h]
\begin{tabular}{cccc}
\hline
\textbf{Model}  & \textbf{MAE ($eV/$atom)} & \textbf{MSE ($eV/$atom)} \\ \hline
%Linear                 & 0.100                       & 0.030                      \\ \hline
CGCNN (Original)               & 0.039                   & -                  \\ \hline
MEGNet (Original)               & 0.028                      & -                       \\ \hline\hline
Linear Model         & 0.302                       & 0.169                     \\ \hline
Random Forest         & 0.100                       & 0.038                     \\ \hline
CGCNN (Ours)               & 0.023                   & 0.003                  \\ \hline
MEGNet (Ours)               & 0.031                      & 0.004                       \\ \hline
\end{tabular}
\caption{Performance on the test set of the various models on prediction of the formation energy per atom using the Materials Project dataset.
}
\label{results_mp_energy}
\end{table}

\begin{table}[h]
\begin{tabular}{cccc}
\hline
\textbf{Model}  & \textbf{MAE ($\mu_B/$atom)} & \textbf{MSE ($\mu_B/$atom)} \\ \hline
Linear                 & 0.100                       & 0.030                      \\ \hline
Random Forest        & 0.043                       & 0.015                      \\ \hline
CGCNN               & 0.052                   & 0.026                  \\ \hline
MEGNet               & 0.052                      & 0.026                       \\ \hline
\end{tabular}
\caption{Performance on the test set of the various models on prediction of the magnetic moment per atom using the Materials Project dataset.
}
\label{results_mp}
\end{table}

\subsection{Predictions on ICSD}
\label{Sec:Discussion}

\begin{figure}[h]
\includegraphics[width=1\columnwidth]{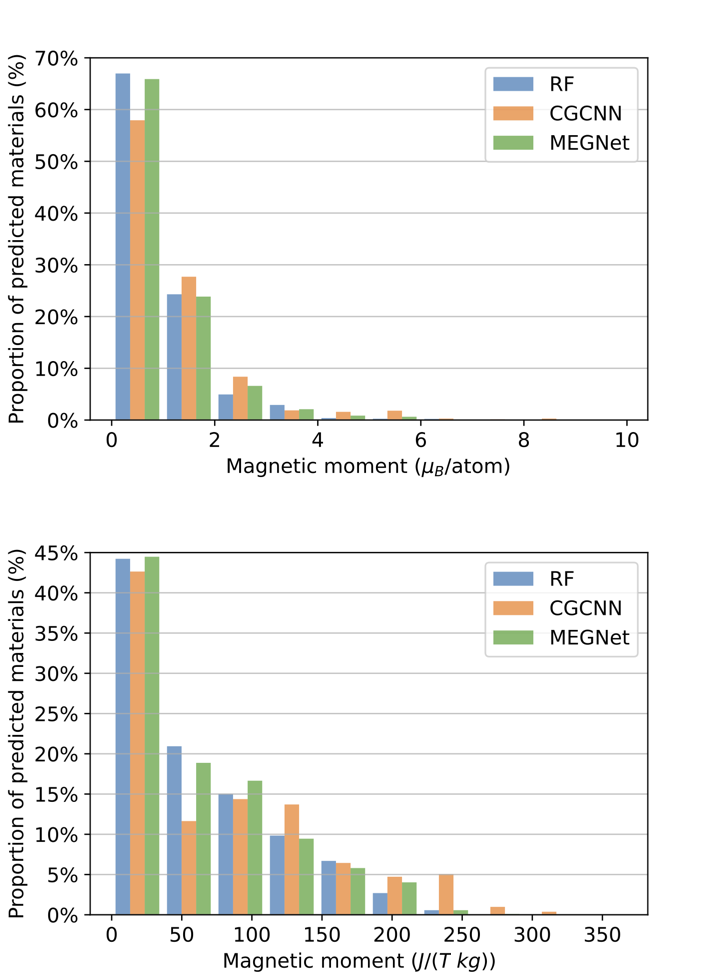}
\caption{\label{fig:predictions_histogramme} Distribution of the predicted magnetic moment for the three models: random forests (RF), CGCNN and MEGNet. Top panel shows the results in units of the Bohr magneton per atom (bin size is one Bohr magneton), bottom panel in units of magnetization per kg.}
\end{figure}

We now apply the trained random forests, CGCNN and MEGNet models on the ICSD dataset. We first start with the disclaimer that the predictions shown below are limited in two ways: (1) by the ferromagnetic bias of the Materials project dataset and (2) by the difficulty of predicting the magnetic moment, as shown in the previous subsection. We use the median and standard deviation of the results from the three models to estimate the magnetization and the error. In Figure \ref{fig:predictions_histogramme}, we illustrate the distribution of the predicted magnetic moments obtained from our three machine learning models. Despite the ferromagnetic bias present in the Materials Project dataset, we note all three models predict a majority of materials with little to no magnetization. More precisely, CGCNN, the random forests and MEGNet predict that $45$\%, $47$\% and $48$\% of the materials have a magnetic moment smaller than $0.5\mu_B$/atom respectively. In Appendix G of the Supplementary Material, we show an extended analysis of the predictions on ICSD.

To better understand the accuracy of our predictions, we sort the results in decreasing order of the predicted magnetization per kilogram. 
Then, we focus on the first and last $150$ materials with the highest and lowest predicted magnetic moment per mass unit from this list. We were left with a total of about $120$ materials after removing duplicate entries (about $60$ materials with a high magnetic moment and $60$ with a zero magnetic moment). To gain an insight on the accuracy of our model, we compare the predicted magnetic properties of these materials to available experimental measurements and report our findings in Table \ref{tab:experimentalOrders}. Notably, we were unable to find experimental reports on the magnetic properties of about $30$ of the materials that were predicted as having a high magnetic moment, indicating that our models could indeed be used for the discovery of new magnetic materials. We report our predictions for the magnetic moment of $15$ of these materials in Table \ref{prediction_mass}. We note that all of these materials include magnetic rare-earths or magnetic transition metals. It is hence plausible that they are indeed ferromagnets with a large magnetic moment.  

 %{\color[red]{Commentaire sur anisotropie et réfrigération magnétique}}

Coming back to Table \ref{tab:experimentalOrders}, we first comment on the results for the materials with high predicted magnetic moments. We find that $17$ ($53$\%) materials from this list are actually reported as ferromagnetic, or ferrimagnetic with a high magnetic moment per mass unit. The remaining materials, accounting for $47$\% of this sample, are found to be mostly antiferromagnetic. This analysis of a subset of our predictions highlights the challenge in predicting the magnetization from models trained of the Materials Project dataset. The ferromagnetic bias in the training set may explain the discrepancy between the predicted ferromagnetic orders and the actual non-ferromagnetic orders observed experimentally. This emphasizes that the discrepancies between high-throughput DFT calculations and experiments can mean that the high performance of a model trained on a DFT dataset does not necessarily translate to accurate predictions when it is applied to an experimental database. We note that this conclusion applies to other predictions, not only to the prediction of the magnetic properties because the magnetic order impacts all the ground-state properties obtained through a DFT calculations.

The comparison of our predictions for small magnetic moments to available experimental data enables a better understanding of the predictive power of our models. Indeed, for these materials, we find that only $3$ materials ($9$\%) from this sample are experimentally found to be ferromagnetic or ferrimagnetic, with the remaining $29$ being antiferromagnetic, non-magnetic, paramagnetic, or diamagnetic, all of which are orders with zero (or negligible) net magnetic moment. Hence, as noted before from Figure \ref{fig:predictions_histogramme}, the ferromagnetic bias present in the training dataset does not preclude the models from accurately predict small magnetic moments. 

\begin{table}[H]
    \centering
  \begin{tabularx}{\linewidth}{ZZZ}
  \hline
  \textbf{Magnetic order reported from experiments} & \textbf{Number of materials (highest magnetic moments)} & \textbf{Number of materials (lowest magnetic moments)} \\
    \hline
  FM/FiM & $17$ ($53$\%) \cite{brown_gd,FM_2,FM_3,FM_4,FM_5,FM_6,FM_7,FM_8,FM_9,FM_10,FM_11,FM_12,FM_13,FM_14,FM_15,FM_16,FM_17} & $3$ ($9$\%)\cite{lFM_1,lFM_2,lFM_3}\\
  AFM & $12$ ($38$\%) \cite{AFM_1,AFM_2,AFM_3,AFM_4,AFM_5,FM_15,AFM_7,AFM_8,AFM_9,AFM_10,AFM_11,AFM_12} & $7$ ($22$\%) \cite{lAFM_1,lAFM_2,lAFM_3,lAFM_4,lAFM_5,lAFM_6,lAFM_7}\\
  NM/PM & $2$ ($6$\%) \cite{PM_1,PM_2} & $11$ ($34$\%)\cite{CRC,lNM_1,lNM_2,lNM_3,lNM_4,lNM_5,lNM_6,lNM_7,lNM_8}\\
  DM & $0$ ($0$\%) & $10$ ($31$\%)\cite{CRC,DM_1,DM_2,DM_3,DM_4}\\
  Other & $1$ ($3$\%)\cite{O_1} & $1$ ($3$\%)\cite{O_2}\\
    \hline
  \end{tabularx}
  \caption{Distribution of the magnetic orders reported experimentally for $64$ materials present in our predictions on ICSD. Middle (right) column lists the results for the $32$ materials with the highest (lowest) median magnetic moment per kilogram predicted by our models for which we could find experimental data.}
  \label{tab:experimentalOrders}
\end{table}

\begin{table}[H]
 \centering
\begin{tabularx}{\linewidth}{ZZZ}
\hline
\textbf{Chemical formula}& \textbf{Moment per mass} (\si{\joule\per(\tesla\times\kg)} & \textbf{Rare earths}  \\
 \hline
\ch{Fe9O}  & 240 \textpm 40 & No \\
\ch{(Mn2O3)3MnSiO3}  & 230 \textpm 50 & No \\
\ch{EuFe2}  & 220 \textpm 60 & Yes \\
\ch{EuOF}  & 220 \textpm 20 & Yes \\
\ch{Eu2Cu}  & 220 \textpm 40 & Yes \\
\ch{Mn6O(VO4)3(OH)}  & 220 \textpm 80 & No \\
\ch{EuH}  & 220 \textpm 90 & Yes \\
\ch{Fe8C2}  & 210 \textpm 20 & No \\
\ch{Fe8N2}  & 200 \textpm 30 & No \\
\ch{GdAl}  & 200 \textpm 20 & Yes \\
\ch{Eu(CN2)}  & 200 \textpm 60 & Yes \\
\ch{Mn8Si6O24ClH9}  & 200 \textpm 60 & No \\
\ch{Gd6Co2Al}  & 200 \textpm 50 & Yes \\
\ch{Mn8O10Cl3}  & 200 \textpm 40 & No \\
\ch{EuBeGd2O5}  & 200 \textpm 80 & Yes \\
 \hline
\end{tabularx}
\caption{Predictions of the largest magnetic moments per mass on the ICSD database for which no magnetic order reported from experiments could be found. Median values and standard deviations of the predictions of random forests, CGCNN and MEGNet are shown.\label{prediction_mass}}
\end{table}

With regards to magnetic refrigeration, the rotating magnetocaloric effect is a promising recent research direction \cite{PhysRevMaterials.4.114411, doi:10.1063/1.4880818, doi:10.1063/1.4943109, Orendac_2018, Zhang_2015}. While the traditional magnetocaloric effect is due to the change in magnetization of a material under the successive application and removal of an external magnetic field, the rotating magnetocaloric effect is due to a strong magnetic anisotropy. In such anisotropic materials, the magnetization is locked in a particular crystal direction (the so-called ``easy-axis"). In that case, a magnetocaloric effect is seen when the material is rotated in a fixed external magnetic field. Materials that exhibit this rotating magnetocaloric effect must have a strong spin-orbit coupling, which is usually associated to elements with a large atomic number. The materials we show in Table \ref{prediction_mass} that include Eu, Gd and In elements could potentially satisfy this requirement.

\section{Conclusion}
\label{Sec:Conclusion}

To identify materials with a large magnetic moment per kilogram, we have trained random forests and two state-of-the-art deep-learning graph convolutional algorithms, CGCNN and MEGNet, on the Materials Project dataset. Since the discovery of large magnetic moments materials could lead to breakthroughs in magnetocaloric refrigeration, we choose this application as a use case. Magnetic properties in the training set are computed using DFT methods. The three machine-learning methods show comparable accuracy on the test sets. Differences in estimates for the mean average error and mean squared error suggest that the predicted magnetic moment per atom is accurate to better than 0.05 Bohr magneton per atom.

We used the trained models to search for candidate materials with large magnetic moments per kilogram in the ICSD database. That database contains materials that have been synthesized. Table \ref{prediction_mass} lists the most promising materials. These deserve experimental attention.
Our analysis also highlights some of the limitations of working with magnetic properties in the Materials Project. In particular, the ferromagnetic bias affects the screening capability of models trained on this dataset.
In the few cases where we could compare our predictions for ferromagnetic materials with experiment, we found that materials were indeed ferromagnetic 50\% of the time. Since the rate of false negatives is low, we believe our model does not leave out many materials that are ferromagnetic.
Finally, we have shown that magnetization may be intrinsically more difficult to predict than other properties. These findings provide strong arguments to motivate building a large database dedicated to magnetic properties of materials.

\section*{Data Availability}

Data from the Materials Project dataset is freely available using their API. Documentation for this API is available at \url{https://docs.materialsproject.org/open-apis/the-materials-api/}.
ICSD is under a proprietary license.

\section*{Code Availability}

Code for the machine learning models is available from the corresponding author upon reasonable request.

\acknowledgments

{We are grateful to Hong Guo and Peter Kang for early access to iMat and to Chi Chen, Patrick Fournier, Shengrui Wang, Yoshua Bengio, Simon Blackburn and Michel Côté for useful discussions.

This work was supported in part by the Canada First Research Excellence Fund (Institut quantique IQ3-2019 2(C.G.-N.) and IVADO (S.-O.K. and S.V.)), by the Natural Sciences and Engineering Research Council of Canada (NSERC) under grant RGPIN-2019-05312 (A.-M.S.T.), by a Vanier scholarship from NSERC (C.G.-N.), a USRA scholarship from NSERC (B.G.-P.). S.-O. K.'s research is also supported by the DeepMind Scholarship. Calculations were performed on computers provided by the Canadian Foundation for Innovation, the Minist\`ere de l'\'Education des Loisirs et du Sport (Qu\'ebec), Calcul Qu\'ebec, and Compute Canada.

}

\section*{Author Contributions}
All authors contributed to designing the study and analyzing results.
S.-O.K. wrote the code for the deep learning models.
B.G.-P. and M.-A. G. wrote the code for random forests models.
S.-O.K. and S.V. analyzed the datasets.
S.-O.K., B.G.-P, M.-A.G., A.-M.S.T. and C. G.-N. wrote the paper. A.-M.S.T, S.V. and C.G.-N. provided supervision for the project.

\section*{Competing Interests}

The authors declare no competing interests.

\nocite{supplementary-material}
\nocite{vasp}
\nocite{paw}
\nocite{iMat}
\nocite{Magpie}
\nocite{WikiBad}
\nocite{featimportance}

\bibliography{apssamp.bib}% Produces the bibliography via BibTeX.

\newpage
\clearpage

\setcounter{page}{1}
\appendix
\maketitle

\section{Additional details on Materials Project}
\label{sec:app1}

The Materials Project dataset is constructed using the Vienna Ab Initio Simulation Package (VASP) software \cite{vasp} and the GGA functional. The calculations are done using the projector augmented wave pseudopotentials implemented in VASP. This method has the advantage of reducing computation time while retaining precision \cite{paw}. The initial crystal structures used as input for the calculations are obtained from the Inorganic Crystal Structure Database (ICSD), which we describe below. The calculations then allow for structure relaxation. All calculations are performed at $T=0$, meaning that the dataset should contain the ground state properties of materials. No spin-orbit coupling effects are included.

\begin{figure*}
\includegraphics[width=2\columnwidth]{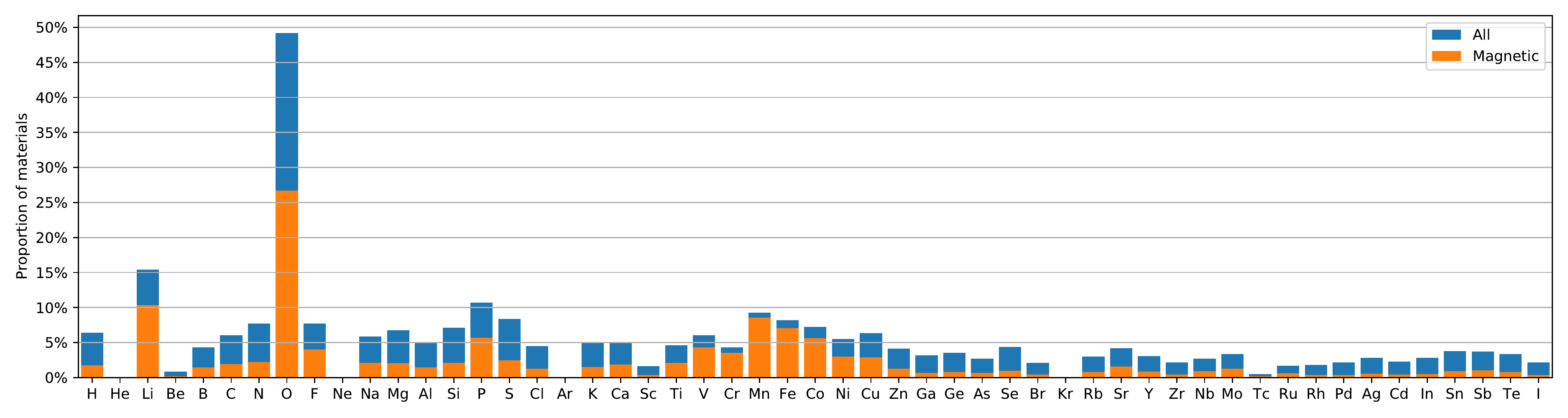}% Here is how to import EPS art

\includegraphics[width=2\columnwidth]{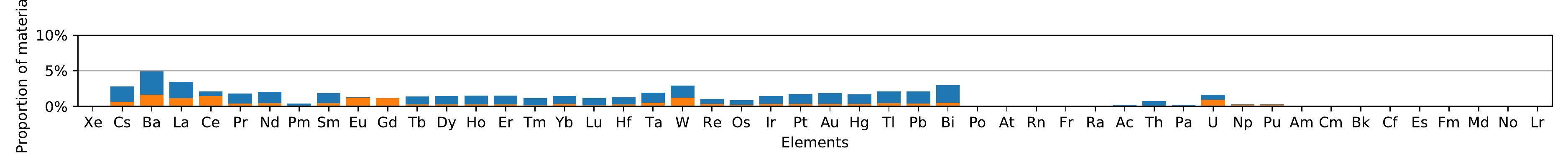}% Here is how to import EPS art
\caption{ Histogram of entries in the Materials Project dataset categorized by the presence (irrespective of stochiometry) of a chemical specie. Blue denotes the distribution over all materials and orange over the subset of magnetic materials.}
\label{fig:element}
\end{figure*}

Figure~\ref{fig:element} shows the distribution of elements in the Materials project dataset, blue indicating the proportion for all materials, and orange the proportion for the subset of magnetic materials. In the latter category, it is not surprising to find transition metals. Oxygen is often present in magnetic materials probably because of the prevalence of perovskite structures that make up 93\% of the lower earth mantle's mass. Surprisingly, Li is also often present in magnetic materials, which may this time reflect the interest of researchers in Li containing materials because of the technological importance of batteries. 
%{\bf Peut-il y avoir une explication à ces dernières remarques? Ce graphique semble-t-il indiquer un biais dans la façon dont materials project est constitué, ex manques-t-il des éléments qu'on s'attendrait à voir plus?}

\section{Hyperparameters}
\label{Sec:Hyperparameters}
For random forests, tuning each hyperparameter individually and in sequence prevents one from detecting possible relations between parameters. Therefore, the ultimate set of hyperparameters was found using random search.

\begin{table}[h]
\begin{tabular}{lr}
\hline
\textbf{Hyperparameter} & \textbf{Value} \\ \hline
Number of descriptors & 98 \\ 
Bootstrap  & None \\ 
Available features per split  & 32.5\% \\ 
Minimum samples for a split   & 4 \\ 
Minimum samples per leaf & 2 \\ 
Minimum depth & None \\ \hline
\end{tabular}
\caption{Value of hyperparameters used for the random forest.}
\label{table:hyperparams}
\end{table}

For neural network methods, we used the same hyperparameters for CGCNN and MEGNet for convenience reasons as well as to avoid excessive hyperparameter tuning. General  hyperparameters were optimized through grid search. Other values were selected in accordance with the original architectures or by convenience given the available hardware.

 \begin{table}[H]
 \centering
 \begin{tabular}{c|lr}
 \hline
 & \multicolumn{1}{c}{\textbf{Hyperparameter}} & \multicolumn{1}{c}{\textbf{Value}}\\
 \hline
 \parbox[t]{3mm}{\multirow{4}{*}{\rotatebox[origin=c]{90}{\footnotesize General}}} & Starting learning rate & 0.001\\
 & Patience & 25\\
 & Learning rate factor & 0.5\\
 & Batch size & 1024\\
 \hline
 \parbox[t]{3mm}{\multirow{4}{*}{\rotatebox[origin=c]{90}{\footnotesize CGCNN}}} 
& Site embedding size & 100\\
& Bond embedding size & 20\\
& Bond Gaussian basis size & 10\\
& Maximum bond distance & 5.2\\
& Hidden layer size & 100\\
& Number of convolution layers & 2\\
& Number of output layers & 3\\
 \hline
 \parbox[t]{3mm}{\multirow{4}{*}{\rotatebox[origin=c]{90}{\footnotesize MEGNet}}} 
& Site embedding size & 36\\
& State embedding size & 36\\
& Bond embedding size & 36\\
& Bond Gaussian basis size & 100\\
& Maximum bond distance & 5.2\\
& Number of MEGNet layers & 3\\
& MEGNet layer hidden size & 64\\
& Number of output layers & 3\\
& Output layer 1 hidden size & 32\\
& Output layer 2 hidden size & 16 \\
 \hline
 \end{tabular}
\caption{Value of hyperparameters used for neural network models.}
\label{table:hyperparams}
 \end{table}

\section{Training set metrics}
\label{sec:train_metrics}

In Table \ref{train_mse}, we report the MSE on the training set for the different method. Small values for the loss function on the training set compared to the test set is evidence of overfitting.

\begin{table}[h]
\begin{tabular}{ccc}
\hline
\textbf{Model}  &  \textbf{MSE ($\mu_B/$atom)} \\ \hline
Random Forest             & 0.015                      \\ \hline
CGCNN                   & 0.0010                  \\ \hline
MEGNet               & 0.0014              \\ \hline
\end{tabular}
\caption{Performance on the training set of the various models on prediction of the magnetic moment per atom using the Materials Project dataset.
}
\label{train_mse}
\end{table}

\section{Random Forest Descriptors}
\label{Sec:RFdecriptors}

Most the descriptors designed for the random forest method are built from the chemical formula of the materials only. Given a material, each atom in the chemical formula have properties specific to themselves and independent of the material. We call these properties \textit{atomic properties}. A list of all atomic properties is given in Table \ref{tab:atomic_properties}. Some of these properties, such as the \textit{atomic number}, the \textit{atomic mass} or the \textit{valence electrons}, are specific to the isolated atom, while other properties, such as the \textit{cohesive energy}, the \textit{density} or the \textit{melting temperature}, are specific to pure materials made of this kind of atom.

\begin{table}[H]
\centering
\begin{tabular}{ll}
\hline
\multicolumn{2}{c}{\textbf{Atomic properties}}                          \\ \hline
Atomic number                         & Mendeleev number              \\ \hline
Column                                & Row                           \\ \hline
Ground state mag. moment              & Mass                          \\ \hline
Electronegativity                     & First ionization energy       \\ \hline
Covalent radius                       & Cohesive energy               \\ \hline
Electron affinity                     & Electron binding energy       \\ \hline
Polarizability                        & Density                       \\ \hline
Boiling temperature                   & Melting temperature           \\ \hline
Volume per atom ($\sqrt[3]{density}$) & Atomic volume (mendeleev)     \\ \hline
Valence electrons (s shell)           & Unfilled electrons (s shell)   \\ \hline
Valence electrons (p shell)           & Unfilled electrons (p shell)   \\ \hline
Valence electrons (d shell)           & Unfilled electrons (d shell)   \\ \hline
Valence electrons (f shell)           & Unfilled electrons (f shell)   \\ \hline
Unfilled electrons (d \& f shells)\,\, & Valence electrons (Magpie)    \\ \hline
Valence electrons (iMat)              & Valence electrons (Wiki.)     \\ \hline
Multiplicity in chemical formula       &                               \\ \hline
\end{tabular}
\caption{List of atomic properties. The \textit{valence electrons} atomic property differed from one source to another (iMat\cite{iMat}, Magpie\cite{Magpie} and Wikipedia\cite{WikiBad}). All three were used. Given a chemical formula A$_x$B$_y$C$_z$, the \textit{multiplicity in the chemical formula} atomic property is $x$ for atom species A, $y$ for atom species B, and $z$ for atom species C.}
\label{tab:atomic_properties}
\end{table}

Then, different operations on these properties can be applied to each chemical formula to build scalar descriptors. All operations used in our method are shown in Table \ref{tab:descriptor_operations}. Thus, given 14 operations and 31 atomic properties, $14\times 31 = 434$ descriptors can be built for each material.

\begin{table}[H]
\centering
\begin{tabular}{ll}
\hline
\multicolumn{2}{c}{\textbf{Operations}}                    \\ \hline
\multicolumn{2}{l}{Maximum}                                      \\ \hline
\multicolumn{2}{l}{Minimum}                                      \\ \hline
\multicolumn{2}{l}{Maximum difference}                           \\ \hline
\multicolumn{2}{l}{Minimum difference}                           \\ \hline
\multicolumn{2}{l}{Weighed average}                              \\ \hline
\multicolumn{2}{l}{Standard deviation}                           \\ \hline
\multicolumn{2}{l}{Average among magnetic atoms (L1, L2)}        \\ \hline
\multicolumn{2}{l}{Weighed average among magnetic atoms (L1, L2, L3)}    \\ \hline
\multicolumn{2}{l}{Maximum among magnetic atoms (L1)}            \\ \hline
\multicolumn{2}{l}{Minimum among magnetic atoms (L1)}            \\ \hline
\multicolumn{2}{l}{Standard deviation among magnetic atoms (L2)} \\ \hline
\end{tabular}
\caption{List of operations that can be associated with each atomic property of Table \ref{tab:atomic_properties} to yield 434 different descriptors. There are two different averages used, \textit{average} and \textit{weighed average}. The \textit{weighed average} is an average that accounts for the occurrence of each atom in the chemical formula, while the \textit{average} computes the average between each atomic species in the chemical formula, independently of how many of each atom there is in the chemical formula.}
\label{tab:descriptor_operations}
\end{table}

Below are the three subsets of elements, L1, L2, and L3 that are involved in some of the descriptors in the table above. In the computation of these descriptors, all elements that are not part of the list specified in parenthesis are attributed a weight zero. These lists help build descriptors that characterise the atoms that are more likely to cause magnetism in a material. Whether an element will induce magnetism in a material is highly dependent on the crystal structure. These lists are thus heuristic. The first list (L1) is the most general. It contains all elements for which one or more of its oxidation states allow magnetism. However, after analyzing the magnetic materials part of the Materials Project database, we notice some elements in L1 nearly never induced magnetism by themselves. In other words, the materials including these elements and no other element from L1 would rarely be magnetic. The set L2 results from the removal of these elements from L1. The set L3, which is the most uncluttered, was also obtained from L1 by removing elements that seemed to be poor inducers of magnetism.\\  

\noindent \textbf{Elements included in the set L1}: Sc, Ti, V, Cr, Mn, Fe, Co, Ni, Cu, Zr, Nb, Mo, Tc, Ru, Rh, Pd, Ag, Ce, Nd, Pm, Sm, Eu, Gd, Tb, Dy, Ho, Er, Tm, Yb, W, Re, Os, Ir, Pt, Au, U, Np, Pu\\

\noindent \textbf{Elements included in the set L2}: Ti, V, Cr, Mn, Fe, Co, Ni, Cu, Nb, Mo, Tc, Ru, Rh, Pd, Ce, Eu, Gd, Tb, W, Re, Os, Ir, U, Np, Pu\\

\noindent \textbf{Elements included in the set L3}: V, Cr, Mn, Fe, Co, Ni, Cu, Ru, Ce, Eu, Gd, U, Np, Pu\\

Descriptors built outside of the \textit{operation} $\times$ \textit{atomic property} scheme are listed in Table \ref{tab:other_descriptors}. Considering that each list (L1, L2, L3), when specified, represents a different descriptor and the fact that there are 7 different crystal systems, we see that there are 29 other descriptors, and 463 descriptors in total. 

\begin{table}[H]
\centering
\begin{tabular}{l}
\hline
\multicolumn{1}{c}{\textbf{Other descriptors}}   \\ \hline
Rate of mag. atoms (L1, L2)                  \\ \hline
Number of species                            \\ \hline
Number of different mag. atoms (L1)          \\ \hline
Fraction of s shell valence electrons        \\ \hline
Fraction of p shell valence electrons        \\ \hline
Fraction of d shell valence electrons        \\ \hline
Fraction of f shell valence electrons        \\ \hline
Crystal system (one-hot encoding)            \\ \hline
Material density                             \\ \hline
Number of sites                              \\ \hline
Magnetic atoms avg. bond length (L1, L2, L3) \\ \hline
Magnetic atoms avg. bond angle (L1, L2, L3)  \\ \hline
Magnetic atoms std. bond length (L1, L2, L3) \\ \hline
Magnetic atoms std. bond angle (L1, L2, L3)  \\ \hline
\end{tabular}
\caption{Descriptors that differ from the \textit{operation} $\times$ \textit{atomic property} building scheme. The last 7 descriptors of the table are the only ones that come from the actual crystal structure of the materials. Only first neighbors are considered in the computation of the average bond length and average bond angle descriptors.}
\label{tab:other_descriptors}
\end{table}

%Might not be necessary
\section{Random Forest Descriptor Selection}
\label{Sec:Features}

Our feature selection method converges iteratively towards the final set of descriptors. The scheme works by chunking the training data and descriptor space into smaller sets on which forward and backward feature selection methods are then performed. 

\begin{figure}[h]
\centering
\includegraphics[width=1\columnwidth]{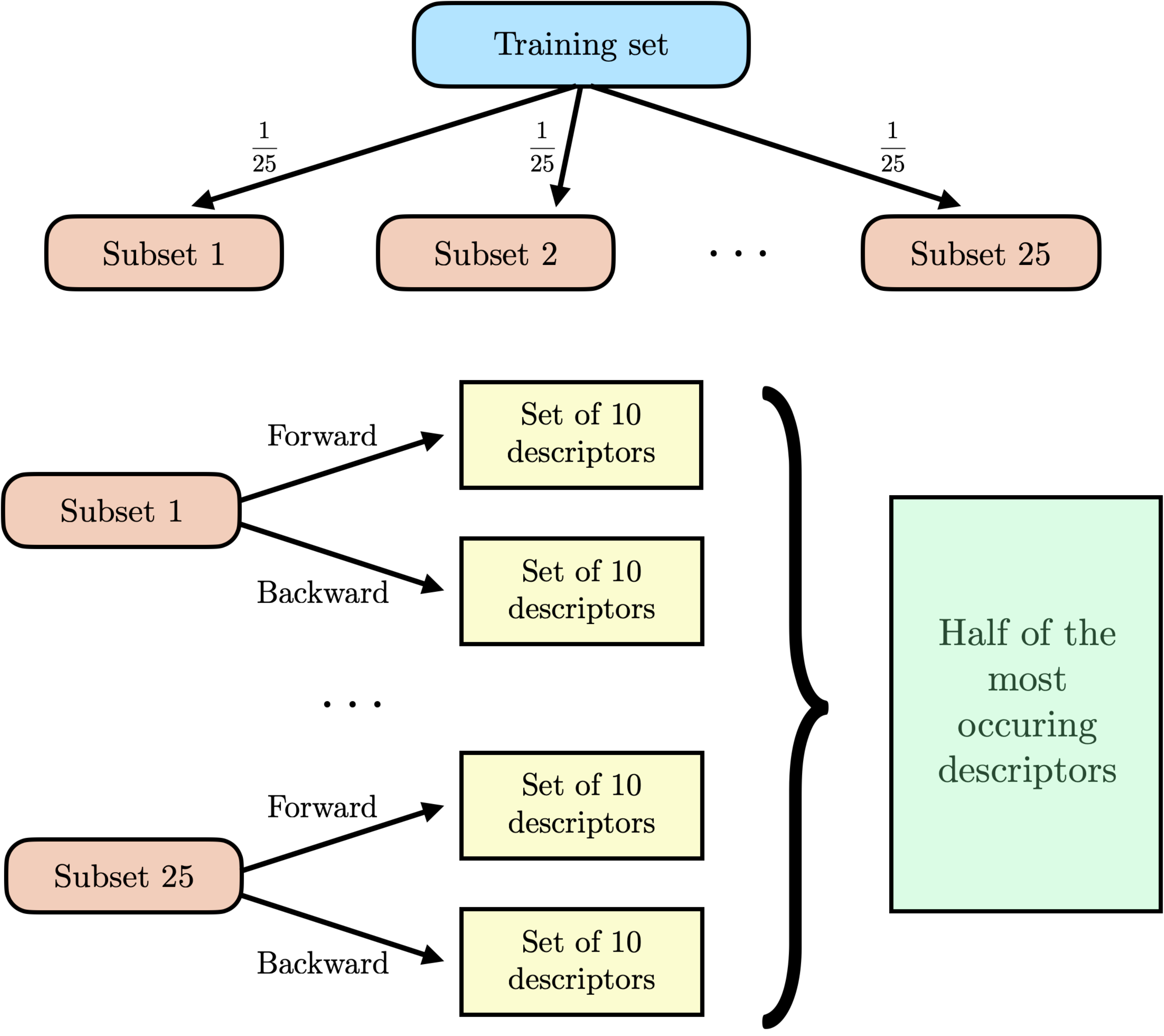}
\caption{Schematic view of one step of the feature selection scheme. 
%At the end of this step, half of the initial predictors are removed. The feature selection task continues by repeating a similar process, but this time by splitting the data in bigger subsets since there are now less predictors.
}
\label{feat_select}
\end{figure}

An example is the best way to illustrate the feature selection method. Suppose one wants to build a random forest regressor from a 100,000-entries training set using only a handful of features from a very large descriptor space. Performing a regular forward or backward feature selection on the whole training set to obtain an optimal set of descriptors would require an unreasonable amount of computational resources and time. To reduce the computational time, one splits its training dataset in 25 equal parts, which results in chunks of data of 4000 materials each. One then performs a forward and a backward feature selection on every chunk of data for a total of 50 calculations. Since the dataset size is much smaller, the computation time is significantly reduced. One then counts the occurrence of each descriptor in the final set of 10 most important descriptors that are selected by each of the 50 calculations and removes half of the descriptors based on their low or null occurrence. A schematic view of this process is shown in Figure \ref{feat_select}. The same operation can be repeated on the new descriptor space to remove even more descriptors and get closer to an optimal final set of descriptors. However, this time, the dataset can be split in fewer chunks since the are less descriptors to choose from. Note that the choices made in this example, such as the number of chunks and the number of descriptors removed at each step, are arbitrary and chosen for convenience.

%\section{Random forest descriptors}

Out of the 463 available descriptors obtained following the method described in Appendix \ref{Sec:RFdecriptors}, this descriptor selection scheme allowed us to select a subset of 98 relevant descriptors. Some of the selected descriptors are still more relevant for the task than others. The importance of each descriptor in determining the magnetic moment is estimated from the mean impurity decrease it provides when selected for a split. The results are shown in Appendix \ref{Sec:FeatureImportance}.  %Clearly, most descriptors have an almost equal and very small importance. Let us focus on the first four descriptors.

The first 16 descriptor account for nearly 75\% of the total importance score. However, the remaining 82 descriptors still play an important role in the decision trees. These descriptors allow finer cuts in the data that are crucial to the algorithm performance. In order to test the relevance of these least important descriptors, we ran the random forest algorithm using only the 16 most important descriptors according to Table \ref{tab:descimportance}, yielding an MAE of 0.072 and an MSE of 0.029, both substantially larger than when the 98 descriptors are used (0.043 and 0.015 respectively in Table \ref{results_mp}).

The descriptor with the highest importance ($22.9\%$) is the ``weighted average of f-shell valence electrons (L3)". This predictor contains information about whether the material contains atoms from L3 that are part of the lanthanide or actinide family or not. The next four most important descriptors are all based on the ``ground state atomic magnetic moment" atomic property. The relevance of these predictors is not surprising. Notice that as a result of the \textit{operation} $\times$ \textit{atomic property} building scheme, the ``Max. at. magnetic moment" and ``Max. among mag. at. magnetic moment" are identical.

\clearpage

\begin{widetext}
\section{Random Forest Descriptor Importance}
\label{Sec:FeatureImportance}

We use the descriptor importance score \cite{featimportance} available in the scikit-learn. The method works by computing how on average each descriptor decreases the impurity when used in a split. The mean impurity decrease is computed for all descriptors and normalized over the entire descriptor set to give the importance score displayed in Table \ref{tab:descimportance}.

\begin{center}

\begin{table}[H]
\small
\resizebox{\textwidth}{!}{
\begin{tabular}{lclc}
\hline
\textbf{Descriptor}                        & \textbf{Importance (\%)} & \textbf{Descriptor}                       & \textbf{Importance (\%)} \\ \hline
Weighted avg. (L3) f shell valence el.       & 22.920                   & Max. among mag at. boiling T            & 0.371                    \\ \hline
Max. at. magnetic moment             & 10.397                   & Weighted avg. (L1) at. column           & 0.336                    \\ \hline
Weighted avg. (L3) at. magnetic moment   & 8.172                    & Std. at. covalent radius           & 0.328                    \\ \hline
Max. among mag at. magnetic moment       & 6.305                    & Std. at. valence el. (Wiki.)              & 0.322                    \\ \hline
Avg. among mag. (L2) at. magnetic moment    & 4.835                    & Min. among mag el. binding energy       & 0.315                    \\ \hline
Weighted avg. (L3) at. volume (mendeleev)     & 3.032                    & Weighted avg. (L3) mendeleev number     & 0.304                    \\ \hline
Weighted avg. (L3) at. valence el. (iMat)      & 2.964                    & Avg. among mag. (L2) at. valence el. (iMat)      & 0.286                    \\ \hline
Weighted avg. (L3) at. d+f unfilled el.      & 2.656                    & Weighted avg. at. valence el. (Magpie)                & 0.285                    \\ \hline
Weighted avg. (L3) at. volume            & 2.479                    & Avg. among mag. (L2) at. density           & 0.254                    \\ \hline
Weighted avg. (L3) at. polarizability    & 2.289                    & Weighted avg. at. valence el. (Wiki.)                  & 0.250                    \\ \hline
Min. among mag at. electron affinity     & 1.862                    & Weighted avg. at. p valence el.                     & 0.237                    \\ \hline
Avg. among mag. (L2) at. electron affinity  & 1.744                    & Std. multiplicity in the chemical formula                     & 0.236                    \\ \hline
Weighted avg. (L3) f shell unfilled el.      & 1.376                    & Min. among mag Z                        & 0.225                    \\ \hline
Weighted avg. (L3) at. unfilled el.   & 1.374                    & Avg. among mag. (L2) at. electronegativity & 0.212                    \\ \hline
Avg. among mag. (L2) at. boiling T          & 1.179                    & Weighted avg. multiplicity in the chemical formula                         & 0.204                    \\ \hline
Weighted avg. (L3) multiplicity in the chemical formula             & 1.154                    & Max. among mag at. melting T            & 0.184                    \\ \hline
Weighted avg. (L3) at. valence el. (Magpie)    & 0.900                    & Max. at. volume                     & 0.181                    \\ \hline
Weighted avg. (L3) at. valence el. (Wiki.)      & 0.865                    & Avg. among mag. (L2) at. column            & 0.173                    \\ \hline
Max. at. cohesive energy             & 0.840                    & Min. at. cohesive energy            & 0.170                    \\ \hline
Weighted avg. (L2) f shell valence el.       & 0.738                    & Max. f shell valence el.                & 0.162                    \\ \hline
Std. at. electron affinity          & 0.724                    & Max. among mag multiplicity in the chemical formula                & 0.146                    \\ \hline
Mag. avg. bond length (L3)                                & 0.694                    & Min. among mag at. valence el. (Magpie)       & 0.142                    \\ \hline
Weighted avg. (L3) at. mass              & 0.691                    & Min. among mag at. mass                 & 0.137                    \\ \hline
Weighted avg. (L2) at. magnetic moment   & 0.687                    & Max. at. melting T                  & 0.131                    \\ \hline
Weighted avg. at. cohesive energy                & 0.648                    & Max. difference multiplicity in the chemical formula              & 0.119                    \\ \hline
Mag. avg. bond length (L2)                                & 0.647                    & Max. at. polarizability             & 0.112                    \\ \hline
Nb. sites                                & 0.634                    & Max. among mag mendeleev number         & 0.110                    \\ \hline
Std. among mag f shell unfilled el.     & 0.565                    & Min. at. boiling T                  & 0.108                    \\ \hline
Weighted avg. (L3) at. electron affinity & 0.563                    & Max. multiplicity in the chemical formula                      & 0.104                    \\ \hline
Mag. std. bond angle (L3)                                & 0.537                    & Max. at. column                     & 0.104                    \\ \hline
Mag. std. bond length (L3)                                & 0.500                    & Max. at. p valence el. el.                  & 0.101                    \\ \hline
Mag. avg. bond length (L1)                                & 0.497                    & Max. d shell valence el.                & 0.086                    \\ \hline
Weighted avg. at. unfilled el.                & 0.484                    & Min. at. electronegativity          & 0.085                    \\ \hline
Weighted avg. f shell unfilled el.                   & 0.474                    & Max. difference at. p valence el.          & 0.083                    \\ \hline
Weighted avg. at. electron affinity              & 0.468                    & Max. at. valence el. (iMat)               & 0.075                    \\ \hline
Weighted avg. at. volume                         & 0.463                    & Max. at. valence el. (Wiki.)               & 0.066                    \\ \hline
Max. difference at. volume              & 0.448                    & Max. at. unfilled el.            & 0.066                    \\ \hline
Weighted avg. (L3) at. column            & 0.433                    & Min. among mag first ionization energy  & 0.065                    \\ \hline
Weighted avg. at. boiling T                      & 0.431                    & Max. difference d shell unfilled el.       & 0.063                    \\ \hline
Std. at. p valence el.                  & 0.420                    & Min. among mag mendeleev number         & 0.062                    \\ \hline
Weighted avg. (L2) multiplicity in the chemical formula             & 0.415                    & Max. among mag at. volume               & 0.060                    \\ \hline
Mag. std. bond angle (L2)                                & 0.413                    & Max. among mag d shell valence el.          & 0.054                    \\ \hline
Mag. std. bond length (L2)                                & 0.398                    & Max. at. d+f unfilled el.               & 0.049                    \\ \hline
Avg. among mag. (L2) at. volume             & 0.397                    & Max. among mag at. d+f unfilled el.         & 0.046                    \\ \hline
Mag. std. bond length (L1)                                & 0.397                    & Avg. among mag. (L2) at. row               & 0.042                    \\ \hline
Weighted avg. (L1) multiplicity in the chemical formula             & 0.389                    & Min. among mag at. electronegativity    & 0.040                    \\ \hline
Weighted avg. at. melting T                      & 0.386                    & Max. f shell unfilled el.               & 0.026                    \\ \hline
Min. el. binding energy              & 0.381                    & Avg. among mag. (L2) f shell unfilled el.      & 0.021                    \\ \hline
Weighted avg. at. density                        & 0.375                    & Max. among mag f shell unfilled el.         & 0.020                    \\ \hline
\end{tabular}}
\caption{List of descriptors ordered by importance score}
\label{tab:descimportance}
\end{table}
\end{center}

\end{widetext}

\begin{figure*}[ht]
         \centering
         \includegraphics[width=0.32\textwidth]{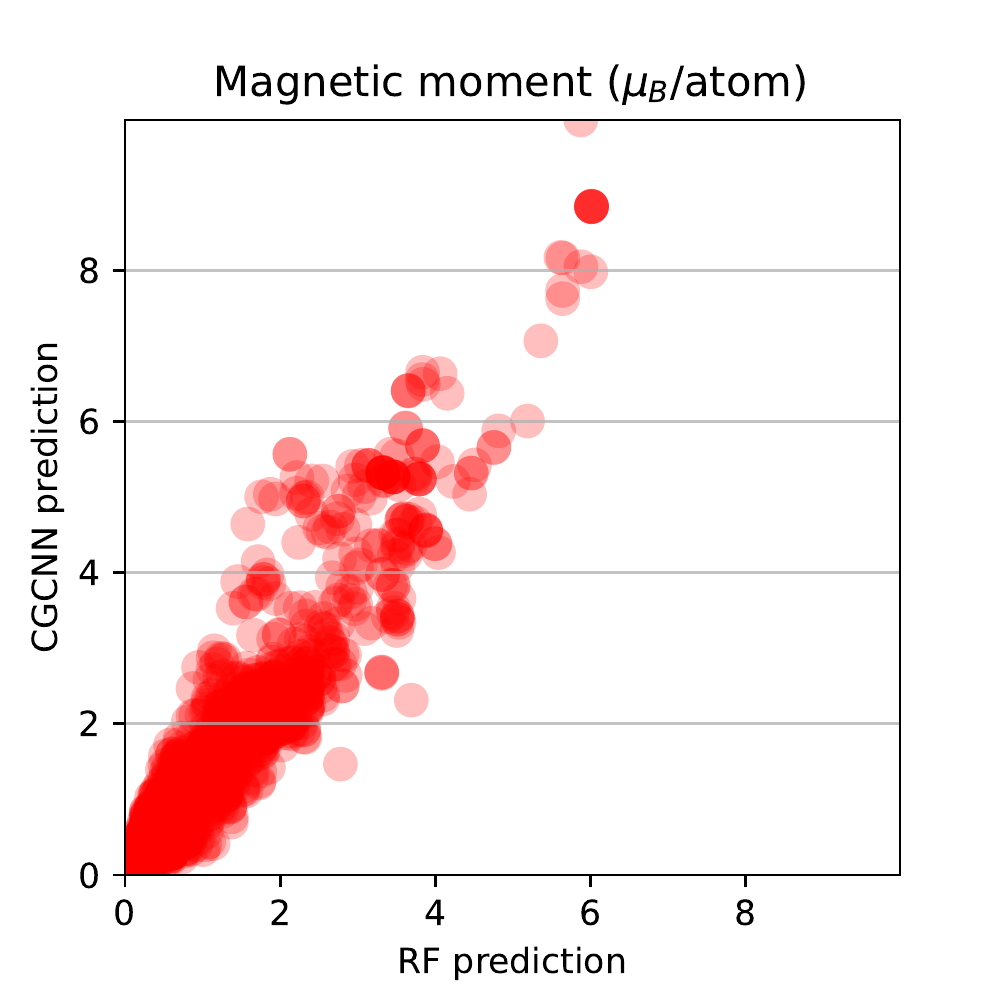}
         \includegraphics[width=0.32\textwidth]{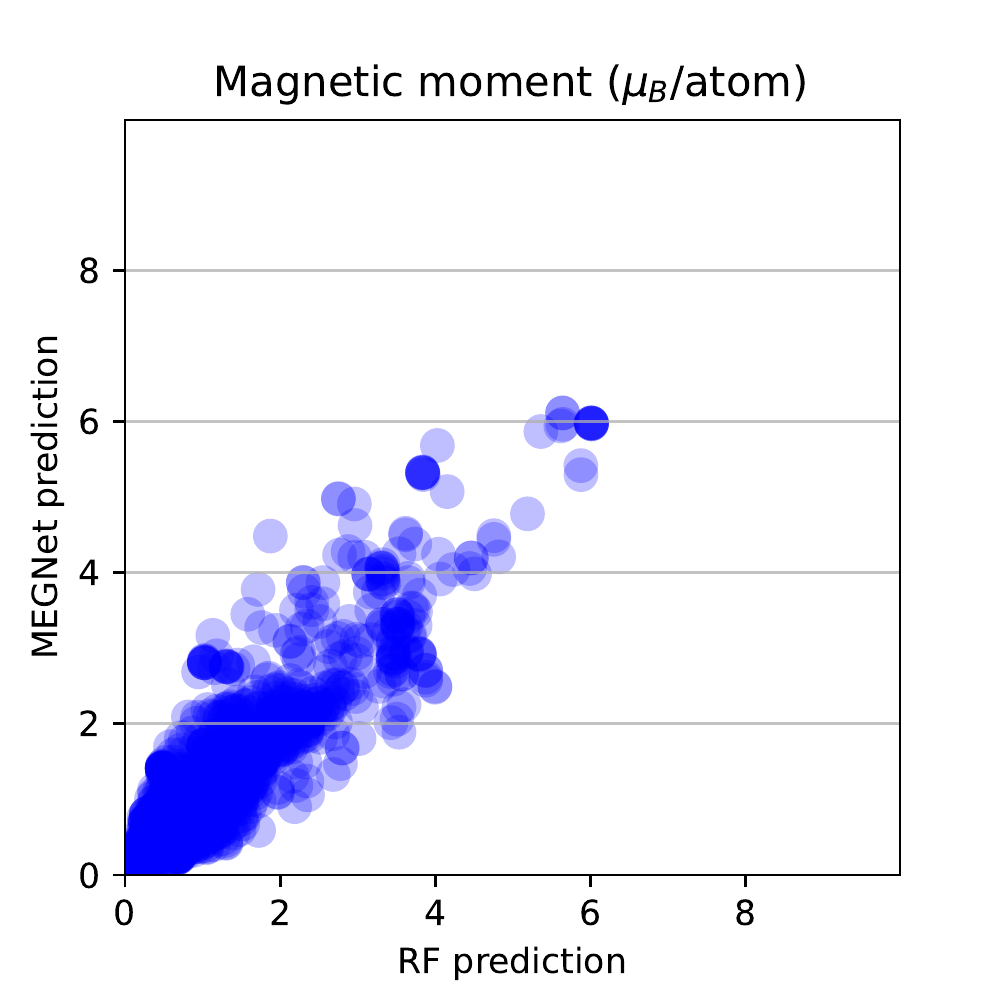}
         \includegraphics[width=0.32\textwidth]{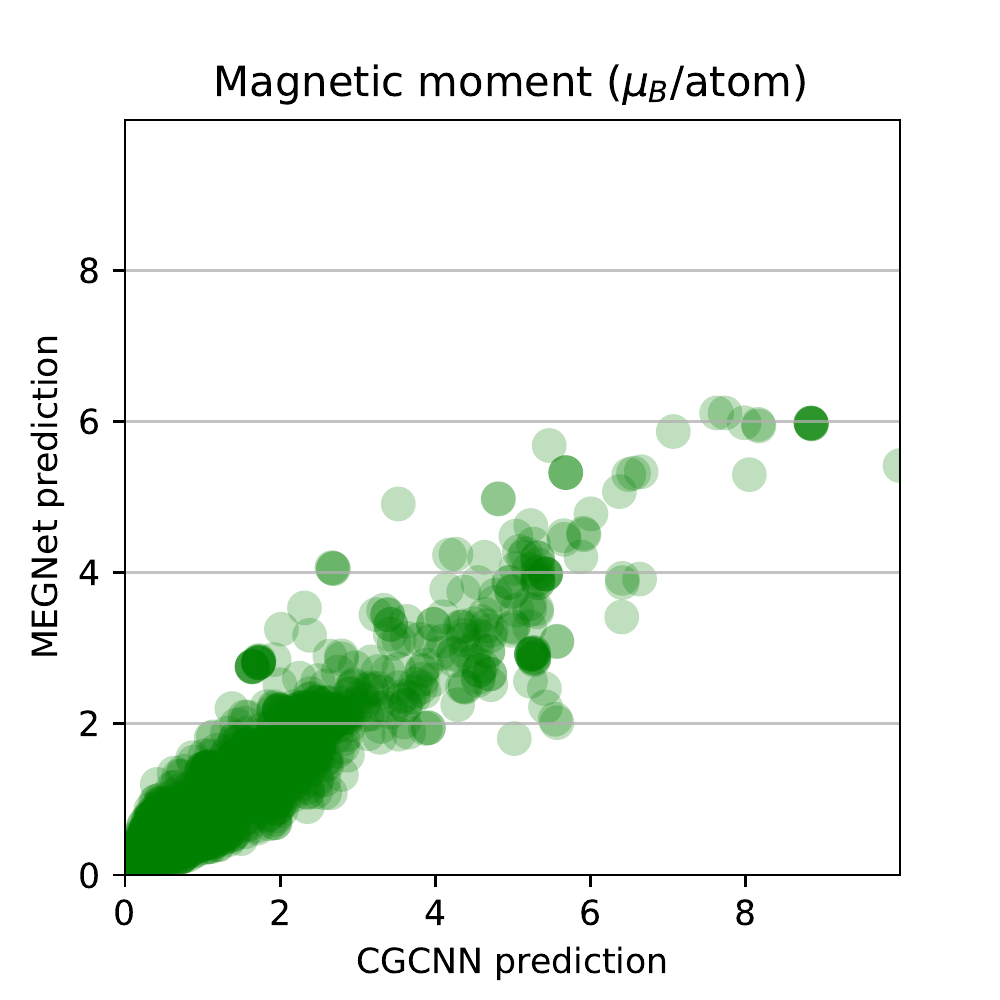}
        \caption{Scatter plots of the predicted magnetic moment per atom as a function of the machine learning models.}
        \label{fig:comp_methods}
\end{figure*}

\begin{figure*}[ht]
         \centering
         \includegraphics[width=0.48\textwidth]{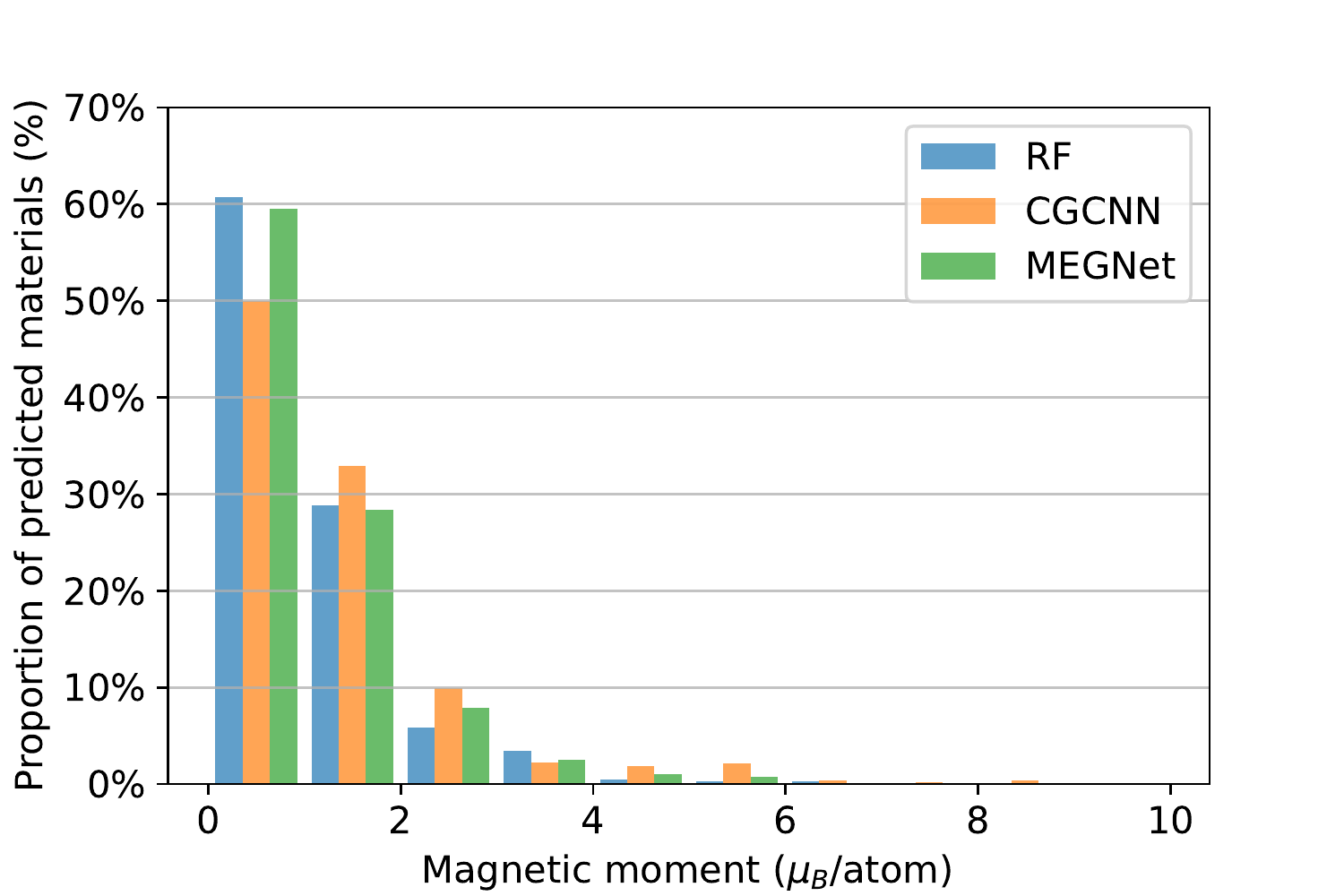}
         \includegraphics[width=0.48\textwidth]{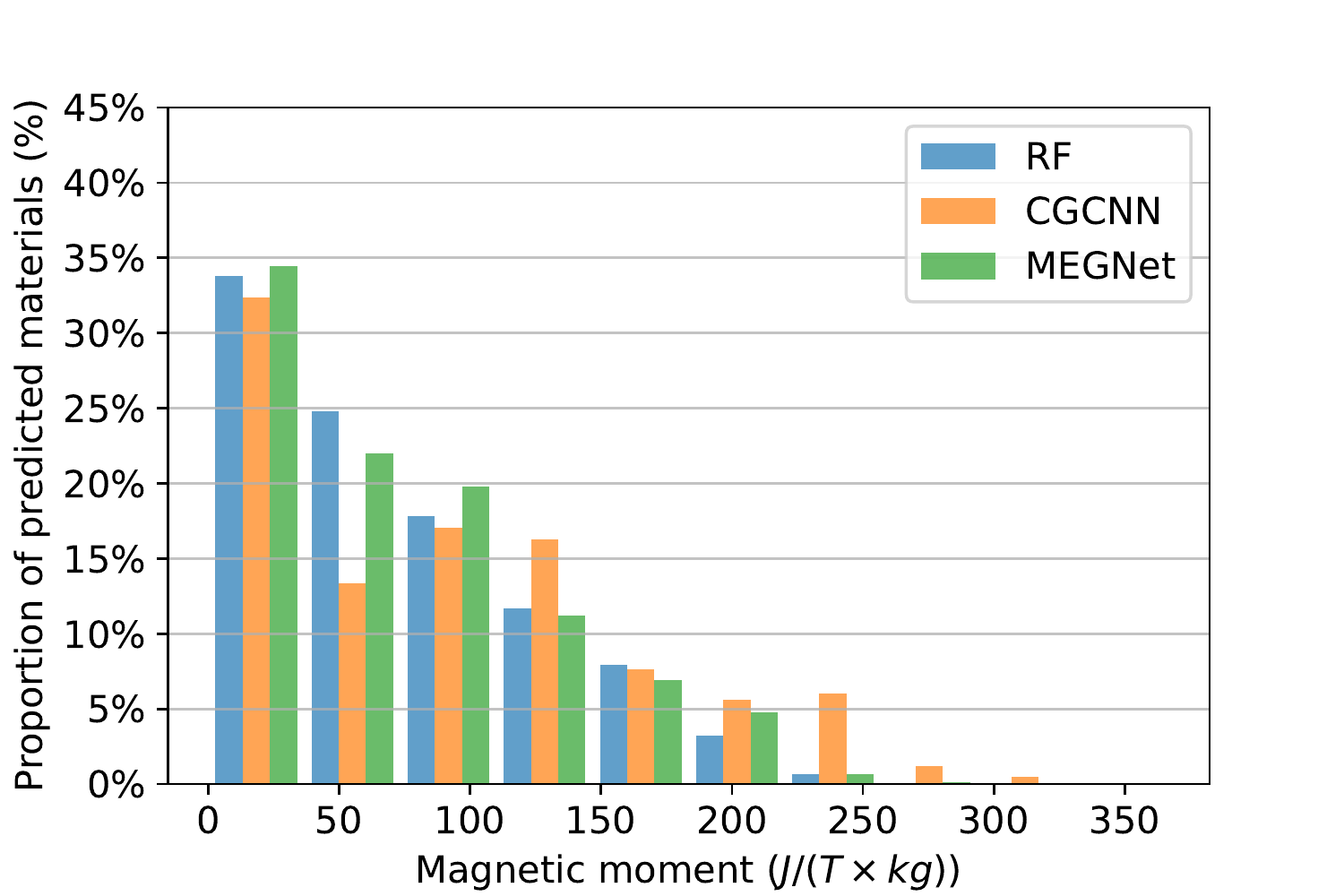}
        \caption{Histograms of the predicted magnetic moments for materials that contain at least one magnetic atom as defined by the L2 list detailed in Appendix \ref{Sec:FeatureImportance}. Left (right) panel shows the magnetic moment per atom (kg).}
        \label{fig:histogrammesL2}
\end{figure*}

\section{More details on the predictions on ICSD}
\label{Sec:detailsResults}

In this Appendix, we offer more analysis of the predictions on the ICSD database. We start by illustrating the distribution of the predictions as a function of the machine learning model in Figure \ref{fig:comp_methods}. These scatter plots yield information similar to that given in the histograms of Figure \ref{fig:predictions_histogramme}. A perfectly linear distribution would indicate that two methods give the exact same predictions for the same materials. The deviation to linearity observed in the plots instead show the degree of variability offered by the different methods. As anticipated from the histograms shown in Figure \ref{fig:predictions_histogramme}, the random forest and MEGNet predictions are closer to each other than they are to the CGCNN predictions.

Next, we produce histograms of the predicted magnetic moments per atom and per mass unit analogous to the ones shown in Figure \ref{fig:predictions_histogramme}, but this time we only plot the data for the materials that contain at least one magnetic element as defined in the L2 list of Appendix \ref{Sec:FeatureImportance}. The results are shown in Figure \ref{fig:histogrammesL2}. The reason we choose to illustrate the results in this way is to investigate whether or not the models simply predict high values of the magnetic moments for materials that contain magnetic atoms, which would render the predictions trivial. However, comparing Figures \ref{fig:histogrammesL2} and \ref{fig:predictions_histogramme}, we note that the distributions of the predicted magnetic moments for the full dataset and for the filtered one are similar. More specifically, between $35$\% and $38$\% of the materials containing at least one element belonging to the L2 list are predicted as having a magnetic moment smaller than $0.5\mu_B$/atom. These proportions are smaller than the ones for the full dataset ($45$\% to $48$\%). This is expected since the elements of L2 are the ones that were found to result in non-zero magnetism in the Materials Project database. However, the high proportion of small predicted magnetic moments for this subset of the ICSD database leads us to believe that the predictions we obtain are more than a trivial classification of materials containing specific elements.

\pagebreak

% \bibliography{bibsup}

\end{document}